\documentclass[10pt,journal]{IEEEtran}
\usepackage{lineno}
\usepackage{lineno}
\usepackage{url}
\usepackage{graphicx}
\usepackage{epstopdf}
\usepackage{subfigure}
\usepackage{amsmath}
\usepackage{comment}
\usepackage{array}
\usepackage{pbox}
\usepackage{multirow}
\usepackage{todonotes}
\usepackage{floatrow}
\newcommand{\RC}[1]{\textcolor{blue}{#1}}

\usepackage{bm,graphicx}
\usepackage{amsmath}
\usepackage[inline, shortlabels]{enumitem}
\DeclareMathOperator*{\argmax}{arg\,max}

\usepackage{booktabs} 
\usepackage{soul}
\usepackage{color, colortbl}
\usepackage{mathtools}
\usepackage{rotating}
\definecolor{Gray}{gray}{0.9}
\definecolor{darkgray}{rgb}{0.66, 0.66, 0.66}

\begin{document}
\title{OntoDSumm : Ontology based Tweet Summarization for Disaster Events}

\author{\IEEEauthorblockN{Piyush Kumar Garg\IEEEauthorrefmark{1}, Roshni Chakraborty\IEEEauthorrefmark{2}, Sourav Kumar Dandapat\IEEEauthorrefmark{1}
}\\
	\IEEEauthorblockA{\IEEEauthorrefmark{1}Deprtment of Computer Science and Engineering\\
	Indian Institute of Technology Patna, India\\
    \IEEEauthorblockA{\IEEEauthorrefmark{2}Department of Computer Science and Engineering\\
    Aalborg University, Aalborg, Denmark\\
		Email: \IEEEauthorrefmark{1}piyush\_2021cs05@iitp.ac.in,
\IEEEauthorrefmark{2}roshnic@cs.aau.dk,
\IEEEauthorrefmark{1}sourav@iitp.ac.in
}}
	}

\IEEEtitleabstractindextext{%

\maketitle

\begin{abstract}

The huge popularity of social media platforms like Twitter attracts a large fraction of users to share real-time information and short situational messages during disasters. A summary of these tweets is required by the government organizations, agencies, and volunteers for efficient and quick disaster response. However, the huge influx of tweets makes it difficult to manually get a precise overview of ongoing events. To handle this challenge, several tweet summarization approaches have been proposed. In most of the existing literature, tweet summarization is broken into a two-step process where in the first step, it categorizes tweets, and in the second step, it chooses representative tweets from each category. There are both supervised as well as unsupervised approaches found in literature to solve the problem of first step. Supervised approaches requires huge amount of labelled data which incurs cost as well as time. On the other hand, unsupervised approaches could not clusters tweet properly due to the overlapping keywords, vocabulary size, lack of understanding of semantic meaning etc.  While, for the second step of summarization, existing approaches applied different ranking methods where those ranking methods are very generic which fail to compute proper importance of a tweet respect to a disaster. Both the problems can be handled far better with proper domain knowledge. In this paper, we exploited already existing domain knowledge by the means of ontology in both the steps and proposed a novel disaster summarization method \textit{OntoDSumm}. We evaluate this proposed method with $4$ state-of-the-art methods using $10$ disaster datasets. Evaluation results reveal that \textit{OntoDSumm} outperforms existing methods by approximately $2-66\%$ in terms of ROUGE-1 F1 score.
\end{abstract}

}
\maketitle

\IEEEdisplaynontitleabstractindextext

\IEEEpeerreviewmaketitle

\section{Introduction} \label{s:intro}

\par Social media platforms, like Twitter, have become extensively popular in the last decade. A large fraction of users use Twitter for a variety of purposes, such as to share opinions and updates, promote products, spread awareness etc. For example, a good number of users share real-time updates during disaster events~\cite{purohit2020ranking, priya2019should, goyal2019multilevel} on social platforms. 
Government and volunteer organizations utilize these updates to ensure effective disaster response. However, due to the high volume and continuous stream of tweets, it becomes challenging for government or volunteer organizations to have a holistic view of the disaster manually. Therefore, an automated summary of these tweets could immensely help these organizations to decide their immediate plan of action. Although there are a plethora of tweet summarization approaches~\cite{zheng2021tweet, chakraborty2019tweet,  narmadha2016survey}, these approaches are not directly applicable for tweet summarization related to disaster events. This is mainly due to the difference in inherent characteristics and features, which vary across domains. Each domain consists of several prominent categories from which representative tweets need to be selected for the summary. The set of categories, as well as the importance of each category, vary across domains. And hence representative tweets from those categories also vary across domains. Prior studies~\cite{imran2015towards, roy2020classification,imran2016twitter} show that tweets related to a disaster comprises of several categories~\footnote{A category consists of information related to a same topic/sub-event of a disaster.}~\cite{imran2015towards}, like \textit{Infrastructure Damage}~\cite{madichetty2021novel, priya2020taqe}, \textit{Victim Needs}~\cite{basu2018automatic, dutt2019utilizing}, \textit{Volunteer Operations}~\cite{basu2019extracting, imran2020using}, \textit{Emotional Response}~\cite{lifang2020effect}, \textit{Affected Population}~\cite{ghosh2018exploitation}, etc. With our experimental result (which will be discussed shortly), it is revealed that the importance of these categories mentioned above varies. 
Therefore, it is required to have the disaster-specific domain knowledge to generate an effective disaster summary.

\par Existing literature on disaster event summary tries to achieve an effective summary by following two major steps \begin{enumerate*} [itemjoin = \quad] \item segregation of the tweets into categories and \item selection of the representative tweets from each category to form a summary~\cite{rudra2015extracting, roy2020classification, rudra2018classifying, nguyen2015tsum4act} \end{enumerate*}. We found both unsupervised as well as supervised approaches in the literature to achieve the first step of summarization. Existing unsupervised approaches, such as graph-based approaches~\cite{dutta2015graph,dutta2019community, dutta2018ensemble} and topic-based approaches~\cite{nguyen2015tsum4act}, utilize the content similarity of the tweets to automatically categorize the tweets into categories by community detection algorithms~\cite{leung2009towards, fortunato2010community, que2015scalable} or Latent Dirichlet Allocation (LDA) based topic modelling approaches~\cite{blei2003latent}. These approaches don’t use domain knowledge and hence, fail to achieve the required performance. At the same time, another set of literature uses a supervised method for the segregation of tweets into categories. Imran et al.~\cite{imran2014aidr}, Rudra et al.~\cite{rudra2016summarizing, rudra2019summarizing}, and Nguyen et al.~\cite{nguyen2022towards} proposed a supervised approach to achieve this objective. However, there is an inherent issue with the supervised method, which needs a labelled dataset which is costly and time-consuming to obtain.

\begin{figure*}
    \centering
    
    \subfigure[]{\includegraphics[width=0.45\textwidth]{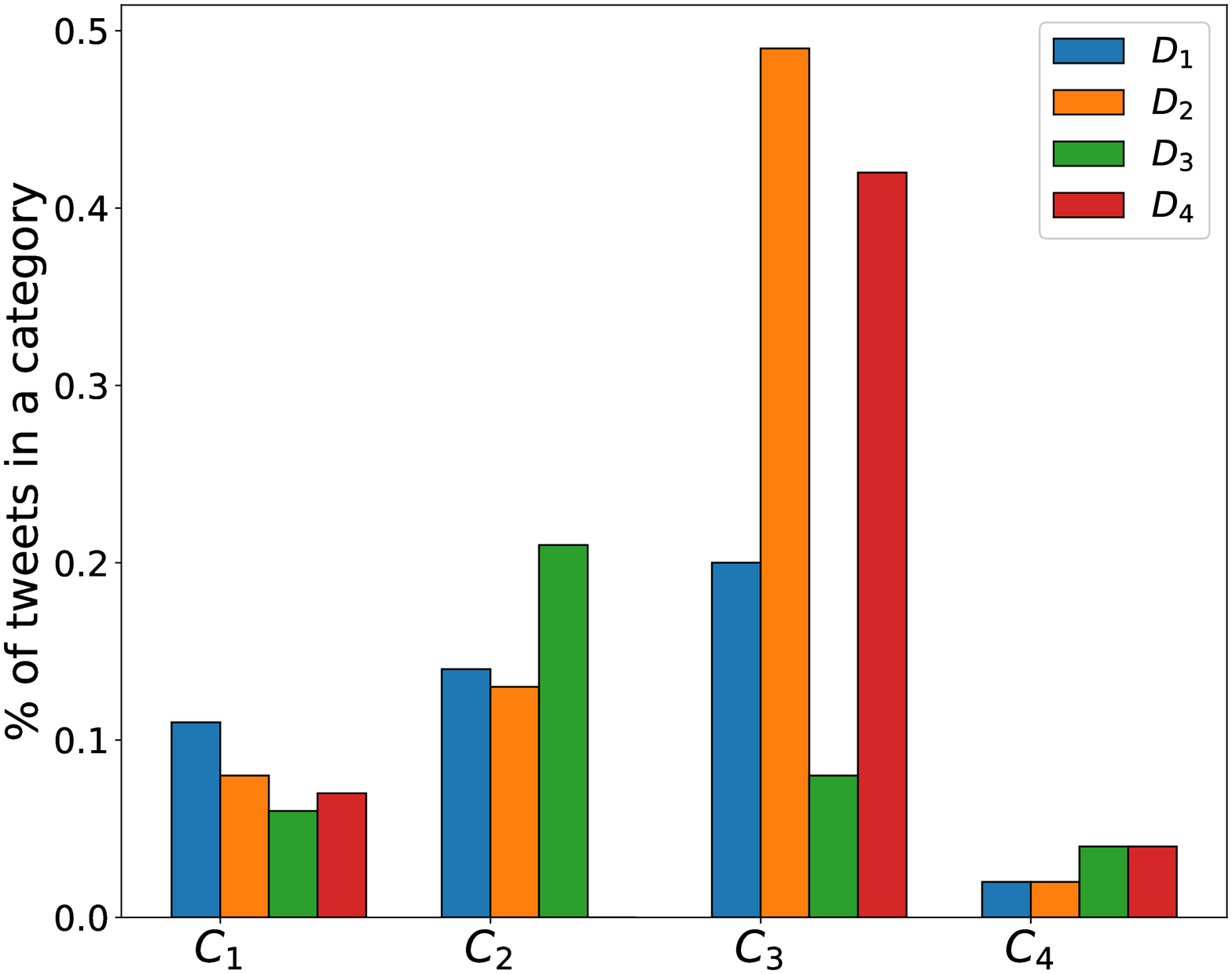} \label{fig:pernumT}}
    \subfigure[]{\includegraphics[width=0.45\textwidth]{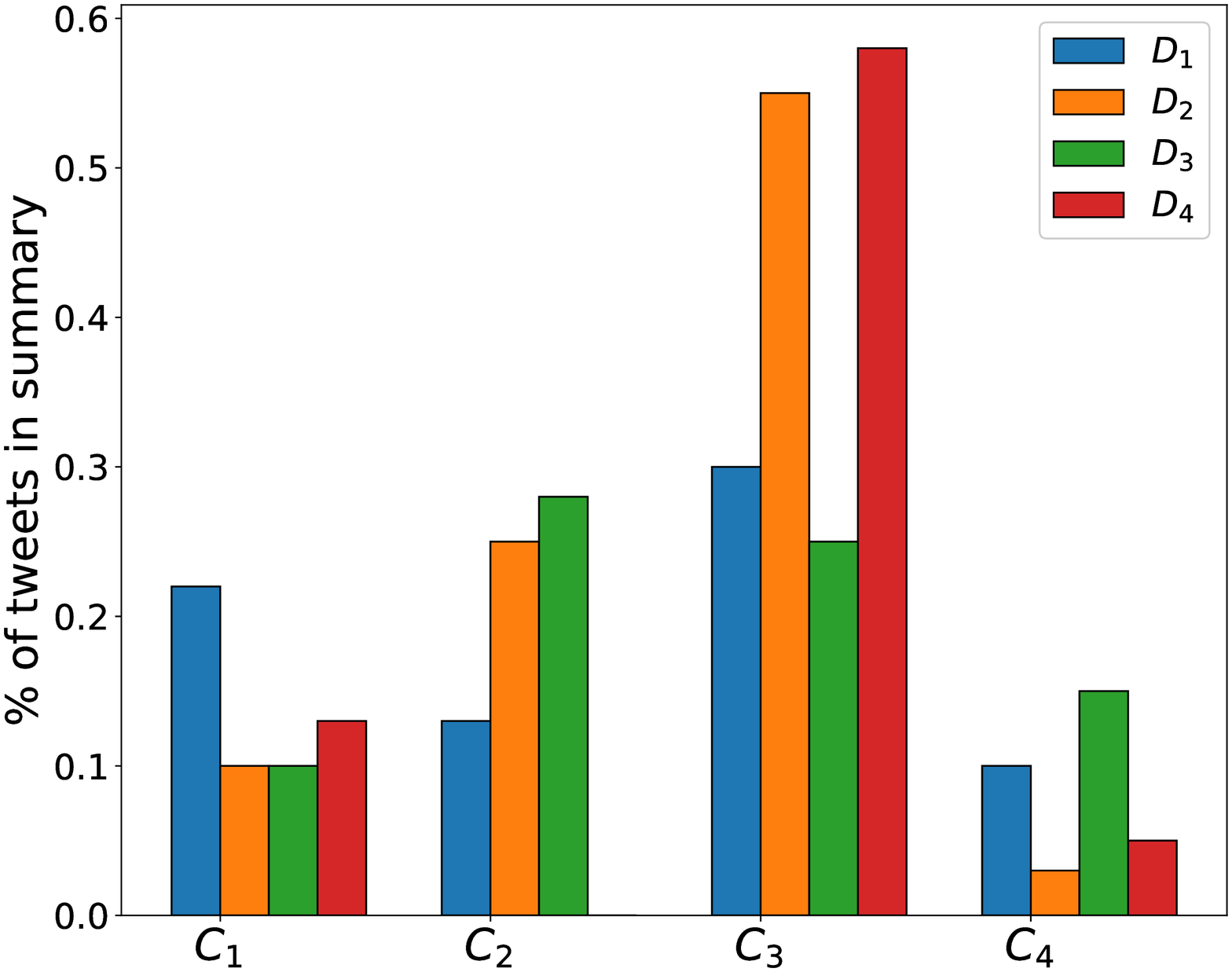} \label{fig:pernumTS}}
\caption{We show the bar-graph of categories \textit{Infrastructure Damage} ($C_1$), \textit{Affected Population} ($C_2$), \textit{Volunteering Support} ($C_3$), and \textit{Early Warning} ($C_4$) of $4$ different disasters, such as \textit{Hurricane Matthew} ($D_1$), \textit{Puebla Mexico Earthquake} ($D_2$), \textit{Pakistan Earthquake} ($D_3$), and \textit{Midwestern U.S. Floods} ($D_4$) for \% of tweets in a category in Figure~\ref{fig:pernumT} and \% of tweets in summary from a category in Figure~\ref{fig:pernumTS}.} 
\label{fig:CatImp}
\end{figure*}

\par For the second step of disaster event summarization, in which representative tweets are selected across categories to form an event summary, existing approaches assign similar importance to all the categories~\cite{rudra2018identifying, dutta2018ensemble, dutta2019community}. For example, Rudra et al. \cite{rudra2018identifying} consider all the categories as equally important and, therefore, select $2$ tweets from each category into the final summary for a disaster, whereas Dutta et al. \cite{dutta2018ensemble} iteratively select a tweet from each category till it reaches to the
desired summary length. Other than event summarization, there are a few summarization approaches which aim to come up with category-specific summaries \cite{nguyen2022towards, rudra2016summarizing, rudra2019summarizing}. Hence, these kinds of summarization approaches do not need to choose representative tweets across categories.


\par Our initial experiments and ground truth summary of a few disaster events reveal that the importance of categories varies within a disaster as well as across disasters, as shown in Figure~\ref{fig:CatImp}. Furthermore, we observe that that while some categories, such as \textit{Affected Population},~\footnote{Information on the number of people missing, displaced, injured or died} \textit{Infrastructure Damage}~\footnote{Damage in buildings, roads, bridges, etc.} are always present irrespective of the disaster, the presence of other categories, such as, \textit{International Aid}~\footnote{Assistance provided by a country or multilateral institutions to the affected country}, and \textit{Aftermath}~\footnote{Consequences of the disaster} depends on the disaster, as shown in Table~\ref{table:Wiki}. 

\begin{table}[ht] 
    \centering 
    \caption{We show the presence of $5$ categories in $6$ disasters, like \textit{Chile Earthquake, 2010} (${\bf D_{1}}$), \textit{Italy Earthquake, 2016} (${\bf D_{2}}$), \textit{India-Pakistan Flood, 2014} (${\bf D_{3}}$), \textit{South Sulawesi Flood, 2019} (${\bf D_{4}}$), \textit{Haiyan Typhoon, 2013} (${\bf D_{5}}$), and \textit{Megi Typhoon, 2010} (${\bf D_{6}}$) based on the information from Wikipedia.} 
    \label{table:Wiki}
    \resizebox{\textwidth}{!}{\begin{tabular} {|>{\centering\arraybackslash}p{0.05\linewidth}|p{0.23\linewidth}|>{\centering\arraybackslash}p{0.05\linewidth}|>{\centering\arraybackslash}p{0.05\linewidth}|>{\centering\arraybackslash}p{0.05\linewidth}|>{\centering\arraybackslash}p{0.05\linewidth}|>{\centering\arraybackslash}p{0.05\linewidth}|>{\centering\arraybackslash}p{0.05\linewidth}|}
        \hline
        {\bf SNo} & {\bf Category} & ${\bf D_{1}}$ & ${\bf D_{2}}$ & ${\bf D_{3}}$ & ${\bf D_{4}}$ & ${\bf D_{5}}$ & ${\bf D_{6}}$ \\ 
        \hline 
        1 & Affected Population   & Yes & Yes & Yes & Yes &  Yes  & Yes \\\hline 
        2 & Infrastructure Damage & Yes & Yes & Yes & Yes &  Yes  & Yes \\\hline 
        3 & Aftermath             & Yes & No  & No  & Yes &  Yes  & Yes \\\hline 
        4 & Donations             & No  & No  & Yes & No  &  Yes  & Yes \\\hline 
        5 & International Aid     & No  & No  & Yes & No  &  Yes  & Yes \\\hline 
    \end{tabular}}
\end{table}

\par In this paper, we propose an ontology-based disaster summarization approach, \textit{OntoDSumm}, which resolves each of these challenges sequentially in three phases. In Phase-I, we propose an ontology-based category identification approach which utilizes the domain knowledge from ontology to map each tweet into a category. Phase-I of \textit{OntoDSumm} is an unsupervised approach; however, it performs far better than other unsupervised approaches as it utilizes the domain knowledge of disaster.  
As previously discussed, the importance of a category varies across disasters, so we propose a method to automatically predict the importance of every category for the given disaster event in Phase-II. Using a metric, “Disaster Similarity Index” (which will be discussed in detail in Section~\ref{s:prop}), it finds a similar disaster and compute category importance from that. Finally, in Phase-III, we propose a modified version of Maximal Marginal Relevance~\cite{carbonell1998use} specifically designed for disaster events, which we refer to as \textbf{D}isaster specific \textbf{M}aximal \textbf{M}arginal \textbf{R}elevance (\textit{DMMR}). The novelty of \textit{DMMR} is that it utilizes ontology knowledge with respect to each category and, therefore, can ensure maximum information coverage of each category in summary. Therefore, by systematic resolution of each of the specific objectives of disaster summarization, \textit{OntoDSumm} can ensure a better performance than the existing state-of-the-art summarization approaches by $2-66$\% in terms of ROUGE-1 F1-scores, $4-77$\% in terms of ROUGE-2 F1-scores, and $3-25$\% in terms of ROUGE-L F1-scores~\cite{lin2004rouge} respectively. We also exhaustively analyze the performance of each Phase of \textit{OntoDSumm} with $9$ different variants of \textit{OntoDSumm} to validate the requirement and role of each Phase of \textit{OntoDSumm}.

\par Rest of the paper is organized as follows. We discuss related works in Section~\ref{s:rworks} and the dataset details in Section~\ref{s:data}. In Section~\ref{s:pstat}, we present problem definition and discuss details of \textit{OntoDSumm} in Section~\ref{s:prop}. We discuss the experiment details in Section~\ref{s:expt}, results in Section~\ref{s:res} and finally, we draw conclusions of \textit{OntoDSumm} in Section~\ref{s:con}.

\section{Related Works} \label{s:rworks}

\par During ongoing important events, a huge serge of tweets is found on Twitter. As a result, many such events are often being reported on social media platforms like Twitter even before any mainstream media~\cite{wold2016twitter} and being considered an important source of information. However, due to noise, duplicate tweets and a humongous amount of tweets, it becomes really difficult to get a precise understanding about ongoing events. Hence, tweet summarization for events got huge attention from the research community. Tweet summarization approaches have been proposed for various domains like sports events~\cite{goyal2019multilevel, huang2018event, gillani2017post}, political events~\cite{panchendrarajan2021emotion, kim2014tweet}, social events~\cite{narmadha2016survey, schinas2015visual}, disasters~\cite{saini2019multiobjective, saini2022microblog}, and news events~\cite{zheng2021tweet, duan2019across, chakraborty2017network}.  

\par Broadly, a disaster summarization approach like any other summarization approach can be categorized as either abstractive~\cite{vitiugin2022cross, nguyen2022towards, lin2021preserve, rudra2016summarize} or extractive summarization~\cite{saini2022microblog, dusart2021issumset, sharma2019going, nguyen2015tsum4act} approaches. As in this paper, our focus is to propose an extractive tweet summarization approach, we restrict ourselves to discuss related literature on the extractive disaster tweet summarization approach.

\par Existing extractive disaster tweet summarization approaches can further be segregated based on their proposed methodologies, such as graph-based~\cite{dutta2018ensemble}, content-based~\cite{sharma2019going}, deep learning-based approaches~\cite{li2021twitter, dusart2021tssubert}, or hybrid of multiple approaches~\cite{saini2020mining}. For example, existing content-based disaster summarization approaches propose different mechanisms to select the tweets into the summary, such as the presence of important words in the tweet~\cite{rudra2015extracting, rudra2018extracting} and coverage of relevant concepts by the tweets. While the presence of important words is measured by the frequency and information content of the words~\cite{rudra2015extracting, rudra2018extracting}, relevant concepts are determined by segregating tweets into either relevant or irrelevant by semi-supervised learning~\cite{chen2015search} or supervised learning~\cite{rudra2018classifying, roy2020classification, madichetty2020detection, madichetty2021neural}. Additionally, deep learning-based approaches have proposed different neural network architectures, such as the disaster-specific BERT model~\cite{liu2019text,dusart2021tssubert} or graph convolutional neural network-based model~\cite{li2021twitter}, to identify the tweets to be selected in summary. However, both content-based and deep learning-based techniques require a huge number of labelled tweets for training which is very difficult to obtain for disaster events. Further, these approaches select a tweet into the summary on the basis of the relevance of that tweet to the disaster and, therefore, do not consider the presence of categories and, correspondingly, the category importance for summary selection and, therefore, fail to ensure the information coverage of each category in summary~\cite{vieweg2014integrating}.


\begin{table*}[ht]
    \centering 
    \caption{We show the details of $10$ disaster datasets, including dataset number, year, number of tweets, type of disaster, and continent.}
    \label{table:dataset}
    \begin{tabular}{|c|c|c|>{\centering\arraybackslash}p{0.09\linewidth}|c|c|c|c|}
        \hline
        {\bf SNo} & {\bf Dataset} & {\bf Year}  & {\bf Number of tweets} & {\bf Type of disaster} & {\bf Continent}\\ \hline 
        1   & $D_1$      & 2012 & 2080 & Man-made & USA  \\\hline
        2   & $D_2$      & 2013 & 2069 & Natural  & Asia \\\hline
        3   & $D_3$      & 2014 & 1461 & Natural  & Asia \\\hline
        4   & $D_4$      & 2013 & 1413 & Man-made & Asia \\\hline
        5   & $D_5$      & 2015 & 1676 & Man-made & Asia \\\hline
        6   & $D_6$      & 2013 & 1409 & Man-made & USA  \\\hline
        7   & $D_7$      & 2016 & 1654 & Natural  & USA  \\\hline
        8   & $D_8$      & 2017 & 2015 & Natural  & USA  \\\hline
        9   & $D_9$      & 2019 & 1958 & Natural  & Asia \\\hline
        10  & $D_{10}$   & 2019 & 1880 & Natural  & USA  \\\hline
    \end{tabular}
\end{table*}

\par In order to incorporate the category-specific information, several research works have proposed graph-based tweet summarization approaches~\cite{dutta2019community, dutta2019summarizing} which initially create the tweet similarity graph with tweets as nodes and an edge as the similarity between a pair of tweets and then group similar tweets together by identifying communities which represent categories. Finally, these approaches select representative tweets from each category based on the length, degree or centrality-based measures to generate the summary~\cite{borgatti2005centrality}. Therefore, these approaches ensure the integration of similar information together by the edge relationships in the graph, implicit identification of categories and further, ensure information coverage and reduction of redundancy by selecting representative tweets from each category. For example, Dutta et al.~\cite{dutta2015graph} propose a community-based approach to identify the different sub-groups as categories from the tweet similarity graph and finally, select representative tweets by centrality-based measures to create a summary. However, these approaches rely on community-based measures to inherently identify the categories of the disaster, which is very challenging due to the high vocabulary overlap across categories in a disaster. Additionally, these approaches consider only content-based similarity to identify the category, which can not ensure handling of the inherent issues of tweets summarization. These approaches also do not consider the difference in importance of categories and their information content across different disasters.

\par Therefore, Rudra et al.~\cite{rudra2016summarizing,rudra2018identifying} use an existing category identification classifier, i.e., AIDR~\cite{imran2014aidr}, to identify the categories and then select representative tweets from each category on the basis of information coverage of a tweet~\cite{rudra2016summarizing} or the presence of disaster information \cite{rudra2018identifying}. However, AIDR requires human intervention for each new disaster event and is applicable only for real-time disaster events. Furthermore, none of these approaches considers the difference in category vocabulary and importance across disasters. Therefore, there is a need to develop a system that can automatically identify the categories of the disaster with minimum human intervention, capture the importance of each category given a disaster and ensure representation of each category on the basis of their specific importance in summary. In this paper, we propose \textit{OntoDSumm} that utilizes disaster-specific knowledge in the form of ontology to identify the category of a tweet which requires no human intervention, followed by importance prediction of a category on the basis of an existing similar disaster and finally, select tweets from each category by a disaster specific selection mechanism to ensure information coverage of each category. We discuss datasets details next. 

\section{Dataset} \label{s:data}
\par In this Section, we discuss the datasets, pre-processing details and gold standard summary.

\subsection{Dataset Details and Pre-processing}  

\par We evaluate the performance of \textit{OntoDSumm} on $10$ disaster datasets which are as follows. An overview of these datasets is shown in Table \ref{table:dataset}.

\begin{enumerate}
    \item \textit{$D_1$}: This dataset is prepared based on the \textit{Sandy Hook Elementary School Shooting}~\footnote{https://en.wikipedia.org/wiki/Sandy\_Hook\_Elementary\_School\_shooting} in which around $26$ people, including $20$ children and $6$ adults were killed in December, $2012$. This dataset is taken from~\cite{dutta2018ensemble}.     
    
    \item \textit{$D_2$}: This dataset is prepared based on the \textit{Uttarakhand Flood}~\footnote{https://en.wikipedia.org/wiki/2013\_North\_India\_floods} which caused dreadful floods and landslides in Uttarakhand, India in June, $2013$. This dataset is also taken from~\cite{dutta2018ensemble}. 
    
    \item \textit{$D_3$}: This dataset is prepared based on the devastating impact of the strong cyclone, \textit{Hagupit Typhoon}~\footnote{https://en.wikipedia.org/wiki/Typhoon\_Hagupit\_(2014)} on Philippines in December, $2014$ which led to the death of around $18$ people and evacuation of $916$ people. This dataset is also taken from~\cite{dutta2018ensemble}.  
    
    \item \textit{$D_4$}: This dataset is prepared based on the \textit{Hyderabad Blast, India}~\footnote{https://en.wikipedia.org/wiki/2013\_Hyderabad\_blasts} in which two consecutive bomb blasts killed $17$ people and injured $119$ people in February, $2013$. This dataset is also taken from~\cite{dutta2018ensemble}.
    
    \item \textit{$D_5$}: This dataset is prepared based on the \textit{Harda Twin Train Derailment,  India}~\footnote{https://en.wikipedia.org/wiki/Harda\_twin\_train\_derailment} in which $31$ people died, and $100$ got injured. The incident was happened in August, $2015$. This dataset is taken from~\cite{rudra2018extracting}.

    \item \textit{$D_6$}: This dataset is prepared based on the \textit{Los Angeles International Airport Shooting}~\footnote{https://en.wikipedia.org/wiki/2013\_Los\_Angeles\_International\_Air- \\port\_shooting} in which around $15$ people were injured and $1$ person was killed. The incident was happened in November, $2013$. This dataset is taken from~\cite{olteanu2015expect}.
    
    \item \textit{$D_7$}: This dataset is prepared based on the devastating impact of the terrible hurricane, \textit{Hurricane Matthew}~\footnote{https://en.wikipedia.org/wiki/Hurricane\_Matthew} on Haiti in October, $2016$ which led to the death of $603$ people and evacuation of $1.5$ million people. This dataset is taken from~\cite{Alam2021humaid}.
    
    \item \textit{$D_8$}: This dataset is prepared based on the \textit{Puebla Mexico Earthquake}~\footnote{https://en.wikipedia.org/wiki/2017\_Puebla\_earthquake} in which $370$ people died, and $6011$ got injured. The incident was happened in September, $2017$. This dataset is taken from~\cite{Alam2021humaid}. 
    
    \item \textit{$D_9$}: This dataset is prepared based on the \textit{Pakistan Earthquake}~\footnote{https://en.wikipedia.org/wiki/2019\_Kashmir\_earthquake} in which $40$ people died, and $850$ got injured. The incident was happened in September, $2019$. This dataset is taken from~\cite{Alam2021humaid}. 
    
    \item \textit{$D_{10}$}: This dataset is prepared based on the \textit{Midwestern U.S. Floods}~\footnote{https://en.wikipedia.org/wiki/2019\_Midwestern\_U.S.\_floods} which caused dreadful floods and massive damages in Midwestern United States in March $2019$ to December $2019$. This dataset is taken from~\cite{Alam2021humaid}. 
\end{enumerate}

\begin{table*}[ht]
    \centering 
    \caption{We show notations and their corresponding description used in \textit{OntoDSumm}.}
    \label{table:notation}
    \begin{tabular}{|c|c|}
        \hline
        {\bf Notation} & {\bf Description}  \\ \hline 
    
        $D$             & Dataset for a disaster event  \\\hline
        $n$             & Number of tweets in $D$ \\\hline
        $m$             & Desired length summary (number of tweets) \\\hline
        $O$             & Existing ontology  \\\hline
        $T$             & Set of tweets in $D$ \\\hline  
        $K$             & Total number of categories \\\hline
        $\mathcal S$     & Generated summary \\ \hline
        $C^i$           & $i^{th}$ category in $D$ \\ \hline
        $T_j$           & $j^{th}$ indexed tweet  \\ \hline
        $I_i$           & Importance of a category $C^i$ \\ \hline
        $OV({C^i})$       & Ontology vocabulary of $C^i$ \\ \hline
        $In(C^i)$       & Aspects present in $C^i$ \\ \hline
        $\alpha, \beta, \lambda$ & Tunable parameters \\ \hline
        $Kw({T}_j)$     & Keywords of $T_j$, such as nouns, verbs, and adjectives  \\ \hline
        $Kw(C^i)$       & Comprise keywords of $C^i$ \\ \hline
    $Div({T}_j,\mathcal S)$ & Diversity if tweet ${T}_j$ added to $\mathcal S$  \\ \hline
    $Sim_1(T_j,OV({C^i}))$ & Contextual similarity of $T_j$ with $O^{C^i}$ \\ \hline
    $Sim_2(T_j, T_l)$    & Cosine similarity between the keywords of $T_j$ and $T_l$ \\ \hline
    $ICov({T}_j,In(C^i))$  & Information coverage provided by tweet, $T_j$ of $C^i$  \\ \hline
    $SemSIM(T_j,C^i)$    & Semantic Similarity score between $T_j$ and $C^i$  \\ \hline 
    $DisSIM(D_x,D_y)$    & Similarity between a pair of disasters, i.e., $D_x$ and $D_y$ \\ \hline  
    $MaxSIM(T_j)$        & Highest Semantic Similarity score among all $C^i$ for $T_j$ \\ \hline
    
    \end{tabular}
\end{table*}

{\textit{Pre-processing and Gold Standard Summary}:} As we consider only the tweet text, we perform pre-processing to remove \textit{URLs}, \textit{usernames}, \textit{emoticons}, \textit{punctuation marks}, and \textit{stop words}. Additionally, as shown by Alam et al.~\cite{alam2018crisismmd}, we observe that the words with length less than $3$ characters do not provide any relevant information specific to disasters, so we remove these words as \textit{noise} from tweet text. We use gold standard summary provided by Dutta et al.~\cite{dutta2018ensemble} for $D_1$-$D_4$ and by Rudra et al.~\cite{rudra2018extracting} for $D_5$. For $D_6$-$D_{10}$, we ask $3$ annotators to prepare a summary of $40$ tweets for each dataset. We follow the procedure by Dutta et al.~\cite{dutta2018ensemble} to combine the individual summaries to prepare the final gold standard summary.

\section{Problem Statement} \label{s:pstat}

\par Given a disaster event, $D$, that comprises of $n$ tweets, ${T}=\{{T}_1,{T}_2,\ldots,{T}_n \}$, we aim to create a summary, $\mathcal{S}$ of $  {T}$. As in most summarization applications, we assume that the length of the summary, $m$, is provided. We previously discussed in Section \ref{s:intro} that a disaster tweet summarization approach must ensure information coverage of all the categories present in ${T}$ where information coverage of a category refers to representation of all the important aspects of that category in $\mathcal{S}$~\cite{yan2011evolutionary}. As there are many different mechanisms, such as topics, keywords or a  combination of both content and context based information, to represent aspects~\cite{rudra2016summarize}, we do not provide any specific method to calculate information coverage and only provide an intuition of information coverage next.
We refer $In(C^i)$ as a measure of aspect/information covered by $i^{th}$ category $C^i$. While $ICov({T}_j,In(C^i))$ measures the number of aspects that are present in $C^i$ which are covered by a tweet ${T}_j$ belongs to category $C^i$. 

\par Additionally, apart from information coverage of all the categories present in ${T}$, the tweets selected in summary, $\mathcal{S}$, must be diverse among each other, i.e., no two tweets selected in $\mathcal{S}$ convey same information~\cite{carbonell1998use,yan2011evolutionary}. Therefore, it is required to select that tweet into $\mathcal{S}$ which maximizes the diversity in $\mathcal{S}$. Existing research works measure diversity by the presence of keywords, aspects, content or contextual information that has not been covered by the tweets already selected in summary~\cite{rudra2015extracting, rudra2018identifying}. We use $Div({T}_j,\mathcal S)$ to measure the diversity provided by selecting ${T}_j$ with respect to the already selected tweets of $\mathcal{S}$. Therefore, to create $\mathcal{S}$, we need to select the tweet, $T^*$ that can maximize both $ICov({T}_j,In(C^i))$ and $Div({T}_j, \mathcal S)$ simultaneously as shown in Equation \ref{eq:probForm1}. 

\begin{equation}
    \begin{aligned}
T^*= \bigcup\limits_{i=1}^{K} \bigcup\limits_{j=1}^{I_i} \max_{{T}_j \in {C}^i}  (\alpha \quad ICov({T}_j,In(C^i)) + \beta \quad Div({T}_j,\mathcal S)) \\
s.t. \quad \sum\limits_{i=1}^{K}{I_i}=m\\
\label{eq:probForm1}
\end{aligned}
\end{equation}

\begin{figure*}
    \centering 
    \includegraphics[width=0.85\textwidth] {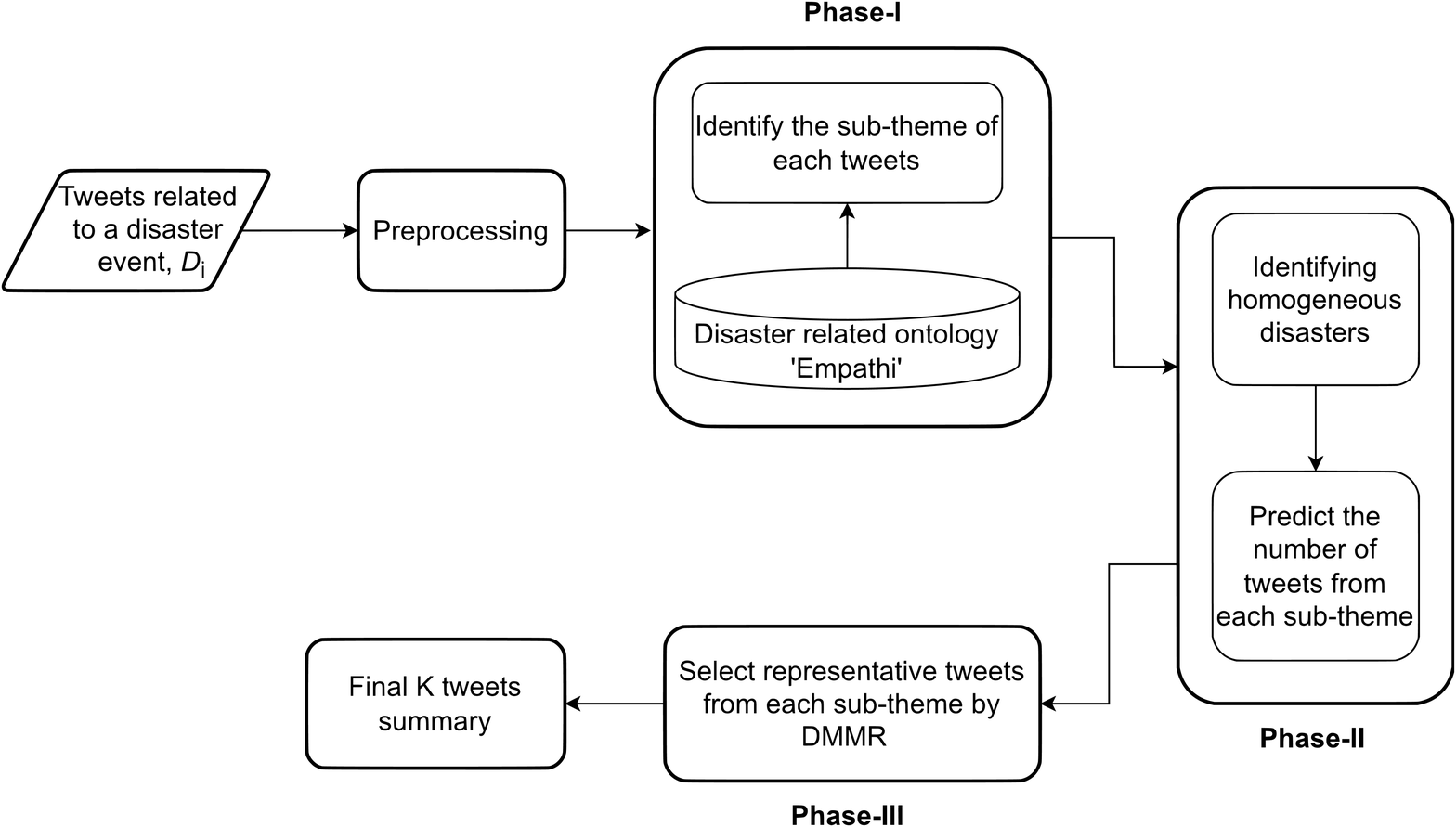}
    \caption{An overview of \textit{OntoDSumm} is shown}
    \label{figure:flowchart}
\end{figure*}

where $I_i$ represents the importance of $i^{th}$ category (which is same as the number of tweets that need to be selected from that category), $\alpha$ and $\beta$ are tunable parameters of information coverage and diversity respectively \cite{carbonell1998use} and $m$ refers to summary length. We refer to the list of categories as ${C}= \{ {C}^1,{C}^2,\ldots, {C}^K\}$ such that there are $K$ categories present in ${T}$. Thus, we intend to select $m$ tweets from ${T}$ such that it maximizes the information coverage present in each category, $In(C^i)$ in ${C}$ and diversity in $\mathcal{S}$ on the basis of $I_i$. We show all the used notations and corresponding descriptions for \textit{OntoDSumm} in Table~\ref{table:notation}.



\section{Proposed Approach}\label{s:prop}
In this Section, we discuss the phases of \textit{OntoDSumm} briefly followed by the details. An overview of \textit{OntoDSumm} is shown in Figure~\ref{figure:flowchart}.

\begin{itemize}
    \item \textit{Identification of the category of a tweet, Phase-I } : We propose an unsupervised approach that utilizes the disaster-specific domain knowledge of an existing ontology, \textit{Empathi}~\cite{gaur2019empathi}, to identify the category of a tweet.
    
    \item \textit{Determination of importance of each category, Phase-II } : We propose a novel score to automatically determine the importance of a category with respect to a given disaster event.
    
    \item \textit{Representative tweets selection from $C^i$, Phase-III } : We propose a Disaster specific Maximal Marginal Relevance to select representative tweets from each category to generate the summary. 
    
\end{itemize}

\subsection{Phase-I} \label{s:phase1} 
In Phase-I, we propose an ontology based pseudo-relevance feedback approach to identify the category of a tweet, which requires no human intervention. Given an existing ontology, $O$, with $K$ categories and their corresponding vocabulary, we propose a \textit{Semantic Similarity score}, $SemSIM(T_j,C^i)$, of $T_j$ with each category, say $C^i$, as shown in Equation \ref{eq:simScore} : 

\begin{align}
   SemSIM(T_j,C^i) = Kw({T}_j) \cap Kw({C}^i)
    \label{eq:simScore}
\end{align}\

where, $Kw({T}_j)$ comprise of the keywords (nouns, verbs, and adjectives) of $T_j$ proposed by Khan et al.~\cite{khan2013multi}, while $Kw({C}^i)$ comprise of the keywords associated with category $C^i$ provided by ontology. On the basis of $SemSIM(T_j,C^i)$, we assign the category of $T_j$ as that with which $T_j$ has the highest \textit{Semantic Similarity score}, $MaxSIM(T_j)$ as shown in Equation  \ref{eq:MaxSim} .
 
\begin{align}
   MaxSIM(T_j) = \underset{i\in K}{\operatorname{\argmax}} (SemSIM(T_j,C^i))
    \label{eq:MaxSim} 
\end{align}

\par We do not propose an ontology for disasters in this paper and use the existing ontology, \textit{Empathi}~\cite{gaur2019empathi}. Although several different disaster-specific ontologies are available~\cite{limbu2012management, Moi2016ontology, sermet2019towards, yahya2020ontology}, we choose \textit{Empathi}~\cite{gaur2019empathi} as it provides the maximum information related to different types of categories compared to others. However, the ontology requires pre-processing. For example, we merge some categories which have similar information, such as three categories, \textit{Infrastructure Damage}, \textit{Broken Bridge}, and \textit{Blocked Road}, are merged as \textit{Broken Bridge} and \textit{Blocked Road} comprises of information which is already covered by definition in \textit{Infrastructure Damage}. Additionally, we observe that the existing keywords of the ontology are not sufficient enough to identify the category of every tweet due to the high vocabulary diversity in tweets. For example, we observe that we could not classify $10-45\%$ of the total tweets with respect to a disaster when we consider only the existing vocabulary of the ontology (as shown in Table~\ref{table:ClasifyTweet}). In order to handle this challenge, we use some disaster related Wikipedia pages to extend the vocabulary of each category. We identify more relevant keywords of each category from the corresponding Wikipedia pages on the basis of similarity between the existing keywords of the ontology category and Wikipedia. Thus using the existing keywords as a query to identify more relevant keywords. We discuss the detailed procedure next.

\begin{table*}[ht]
    \centering\caption{We show the \% of classified tweets using vocabulary of the ontology (i.e., Empathi) and extended vocabulary using disasters Wikipedia pages for $10$ datasets. }
    \label{table:ClasifyTweet}
    \begin{tabular}{|c|>{\centering\arraybackslash}p{0.09\linewidth}|>{\centering\arraybackslash}p{0.1\linewidth}||c|>{\centering\arraybackslash}p{0.09\linewidth}|>{\centering\arraybackslash}p{0.1\linewidth}|}
        \hline
        {\bf Dataset} & {\bf Empathi} & {\bf Extended vocabulary} & {\bf Dataset} & {\bf Empathi} & {\bf Extended vocabulary} \\ \hline 
        $D_1$ & 66.33 & 3.02 & $D_6$     & 65.60 & 5.39 \\\hline
        $D_2$ & 68.16 & 8.72 & $D_7$     & 85.25 & 4.05 \\\hline 
        $D_3$ & 65.31 & 5.51 & $D_8$     & 89.78 & 4.32 \\\hline 
        $D_4$ & 64.64 & 3.77 & $D_9$     & 85.44 & 5.52 \\\hline 
        $D_5$ & 72.54 & 8.36 & $D_{10}$  & 78.25 & 5.54 \\\hline
  \end{tabular} 
\end{table*}

\par 
We, initially, select $20$ Wikipedia pages which belong to different locations and types of disasters. We follow  existing research works~\cite{olteanu2014crisislex, imran2016twitter} to identify the relevant Wikipedia pages. To extend the vocabulary of an ontology category, we first identify all the sentences from Wikipedia pages which consists of at least one keyword of the ontology category. Then we form a potential keyword list corresponding to this ontology category by adding nouns, verbs, and adjectives~\cite{rudra2015extracting} from those sentences. Then we further filter out this potential keyword list based on the frequency and ignored those keywords having a frequency of less than three. Before the final inclusion of these keywords as extended vocabulary, it is given to a human annotator. A keyword is finally added to vocabulary of the ontology category if the annotator finds it relevant. This procedure makes sure extended vocabulary does not have any irrelevant keywords. We follow the same procedure for all ontology categories. Therefore, we now use this extended vocabulary to identify the category of a tweet. On using this extended vocabulary, we observe that around $68.41-94.10$\% of the tweets could be classified with respect to a disaster which is more by around $3-9$\% than without using the extended vocabulary. However, we could not classify around $6-32\%$ of the tweets across disasters even after using the extended vocabulary. We show in Table~\ref{table:ClasifyTweet} the fraction of tweets classified using vocabulary of the ontology and extended vocabulary for $10$ disaster events. We do not consider the tweets whose category could not be determined in Phase-I. This is an important point to be noted that around $3-30\%$ of the total tweets across disasters are irrelevant.


\par In order to understand the performance of Phase-I, we validate by $3$ manual annotators who manually determine the category of 
all the classified tweets based on their understanding of the tweet text and category definition. We consider the ground truth category of a tweet as that category which is selected by the majority of the annotators. We, then, calculate the F1-score on the basis of the ground truth category and automated category as given by \textit{OntoDSumm}. We show our results in Table~\ref{table:F1_score_pass1&2} which indicates almost perfect classification performance by Phase-I of \textit{OntoDSumm} irrespective of the dataset. 

\begin{table}[ht]
    \centering\caption{We show the F1-scores for the classified tweets using vocabulary of the ontology (i.e., Empathi) and extended vocabulary using disasters Wikipedia pages on $5$ datasets, such as $D_6$, $D_7$, $D_8$, $D_9$, and $D_{10}$.}
    \label{table:F1_score_pass1&2}
    \begin{tabular}{|c|>{\centering\arraybackslash}p{0.18\linewidth}|>{\centering\arraybackslash}p{0.2\linewidth}|}
        \hline
        {\bf Dataset} & {\bf Empathi} & {\bf Extended vocabulary}   \\ \cline{2-3}
                      & \bf{F1-score} & \bf{F1-score} \\ \hline
        $D_6$   & 0.997 & 1.000   \\\hline
        $D_7$   & 0.995 & 0.985 \\\hline
        $D_8$   & 0.992 & 0.983 \\\hline
        $D_9$   & 0.989 & 0.991 \\\hline
        $D_{10}$& 0.994 & 0.990 \\\hline
  \end{tabular}
\end{table}

\subsection{Phase-II} \label{s:phase2} Given a disaster, $D_x$, we propose an approach to automatically determine the importance of each category with respect to $D_x$ in this Phase. We need the importance of a category, $I_i$, to determine the number of tweets to be selected from each category in the final summary. Therefore, $I_i$ of $C^i$ represents the total number of tweets to be selected from $C^i$ given $D_x$. Content of the tweets across the categories and the distributions of the tweets across categories vary with disasters. We use a Linear Regression model to predict the number of tweets to be included in the final summary. We train this regression model with a disaster, $D_y$, similar to $D_x$. To identify $D_y$ for a given $D_x$, we propose a metric \textit{Disaster Similarity Index} which we discuss next.

\par \textbf{Disaster Similarity Index :} \label{s:sim}
We propose a \textit{Disaster Similarity Index}, $DisSIM(D_x,D_y)$ to compute the similarity between any pair of disasters, $D_x$ and $D_y$. We define $DisSIM(D_x,D_y)$ as the weighted score of information content of the categories, $Cat_{IC}(D_x,D_y)$ and similarity in probability distribution between categories $Cat_p(D_x,D_y)$. We compute $Cat_p(D_x,D_y)$ as (1-$Cat_{ds}(D_x,D_y)$) where $Cat_{ds}(D_x,D_y)$ is the Jensen Shannon Divergence of two events. $DisSIM(D_x,D_y)$ is calculated as : 

\begin{align}
    \begin{split}
       DisSIM(D_x,D_y) = w_1 * Cat_{IC}(D_x,D_y) + \\w_2 * Cat_{p}(D_x,D_y)
    \end{split}
\end{align}

\begin{align}
   \textrm{s.t.} \quad  w_1 + w_2 =1 \\
   w_1, w_2 \in (0, 1)
    \label{eq:Cweight1} 
\end{align}

where, $w_1$ and $w_2$ are the weights of $Cat_{IC}(D_x,D_y)$ and $Cat_{p}(D_x,D_y)$ respectively and we consider equal weighted of $Cat_{IC}(D_x,D_y)$ and  $Cat_{p}(D_x,D_y)$ as $w_1=w_2=0.5$. We calculate $Cat_{IC}^i(D_x,D_y)$ as the cosine similarity~\cite{nguyen2010cosine} of the most frequent occurring keywords between $D_x$ and $D_y$ for a category, $i$ and $Cat_{IC}(D_x,D_y)$ as the average cosine similarity over all categories as shown in Equation~\ref{eq:Csim}. 
\begin{align}
   Cat_{IC}(D_x,D_y) = \frac{1}{K}\sum_{i=1}^{K} Cat_{IC}^i(D_x,D_y) 
    \label{eq:Csim} 
\end{align}
where $K$ is the total number of categories. We calculate $Cat_{p}(D_x,D_y)$ by the Jensen-Shannon divergence~\footnote{https://en.wikipedia.org/wiki/Jensen\%E2\%80\%93Shannon\_divergence} score of a disaster pair $D_x$ and $D_y$. $Cat_{p}(D_x,D_y)$ measures the similarity of the probability distributions of the categories between $D_x$ and $D_y$. We select $D_q$ as the disaster which has maximum similarity with $D_x$ on the basis of $DisSIM(D_x,D_y)$ where, $y \in {{0,1,\ldots,Q}}$ and $Q$ is the total number of disasters we have. We discuss this experiment and our observations in detail in Section~\ref{s:similar}. 


\subsection{Phase-III} \label{s:phase3}

Several existing tweet summarization approaches \cite{dutta2019summarizing,chakraborty2017network} have proposed Maximal Marginal Relevance, \textit{(MMR)}, to select tweets iteratively such that it  selects that tweet into the summary which provides the maximum relevance with respect to the query and can maintain maximum diversity among the already selected tweets into summary. Therefore, Maximal Marginal Relevance~\textit{(MMR)} inherently do not consider the category relevance which is necessary for disaster summarization. Therefore, we propose \textbf{D}isaster specific  \textbf{M}aximal \textbf{M}arginal \textbf{R}elevance, \textit{DMMR}, for tweet selection from each category. For $C^i$, we iteratively select that tweet which is most relevant to the category and ensures maximum diversity among the already selected tweets into summary till $I_i$ number of tweets are selected. 
We select the tweet, $T^{max}$ that has the maximum score by Equation~\ref{eq:DMMR} as follows : 

\begin{align}
    \begin{split}
        T^{max} = \argmax_{T_j \in C^i}[\lambda*Sim_1(T_j,OV(C^i))-  \\ (1-\lambda)(\max_{T_l \in S} (Sim_2(T_j, T_l)))]
    \end{split}
    \label{eq:DMMR} 
\end{align}

\par where, $S$ is the set of tweets already selected in summary, and $\lambda$ is a hyper-parameter in the interval $[0,1]$ which measures the importance of relevance provided by $T_j$, i.e., $Sim_1(T_j,OV(C^i))$ and diversity of $T_j$ with respect to $S$, i.e., $Sim_2(T_j, T_l)$. We set the $\lambda$ value as $0.5$ by giving equal importance to the relevance and diversity. We explain how we calculate $Sim_1(T_j,OV(C^i))$ and  $Sim_2(T_j, T_l)$ respectively next. In order to capture the relevance of $T_j$ with respect to a category, $Sim_1(T_j,OV(C^i))$, we calculate the contextual similarity of $T_j$ with the ontology vocabulary of $C^i$, i.e., $OV(C^i)$. Therefore, $Sim_1(T_j,OV(C^i))$ ensures consideration of disaster specific category information with respect to $C^i$ as we consider ontology vocabulary. Additionally, contextual similarity based calculation can handle the high vocabulary diversity inherent in tweets as in : 

\begin{align}
   Sim_1(T_j,OV(C^i)) = \sum\limits^{\vbox to 0pt{\hbox{\,\rule{.5pt}{2em}}}}_{v \in Kw(T_j)} CSim(v, OV(C^i))
    \label{eq:SemanticSim_W2V}
\end{align}

\begin{align}
   CSim(v, OV(C^i)) = \max_{k \in Kw(OV(C^i))} (VSim(v, {OV(C^i)}_k))
   \label{eq:CSim}
\end{align}

\begin{align}
   VSim(v, {OV(C^i)}_k)) = \frac{\vec{v} \cdot \vec{{OV(C^i)}_k}}{\vert \vec{v} \vert \ \vert \vec{{OV(C^i)}_k} \vert}
\end{align}

\par where, $Kw(T_j)$ and $Kw(OV(C^i))$ are the set of keywords of $T_j$ and $OV(C^i)$. While $\vec{v}$ and $\vec{{OV(C^i)}_k}$ are the  word embedding of $v$ and ${OV(C^i)}_k$ generated by Word2Vec~\cite{imran2016twitter} respectively. We follow Rudra et al.~\cite{rudra2015extracting} and only consider nouns, verbs and adjectives of $T_j$ as the keywords($v$) of $T_j$. We calculate $Sim_2(T_j, T_l)$ as the cosine similarity~\cite{nguyen2010cosine} between the keywords of $T_j$ and $T_l$ where $T_l$ is the set of tweets that are in $\mathcal S$ from $C^i$. We calculate $Sim_2(T_j, T_l)$ as :

\begin{align}
   Sim_2(T_j, T_l) = \frac{ \vert Kw(T_j) \cap Kw(T_l) \vert}{\sqrt{\vert Kw(T_j) \vert \ \vert Kw(T_l) \vert}}
    \label{eq:consim}
\end{align}

where, $Kw(T_j)$ and $Kw(T_l)$ are the keywords of $T_j$ and $T_l$, respectively. 

\section{Experiments}\label{s:expt}
\par In this Section, we initially discuss the details of the existing research works related to disaster summarization which we use as baselines. Then, we provide a performance comparison of the baselines and \textit{OntoDSumm}.

\begin{table*}[ht]
    \centering 
    \caption{Precision, recall and F1-score of ROUGE-1, ROUGE-2 and ROUGE-L score of \textit{OntoDSumm} and baselines on $5$ datasets, i.e., $D_1-D_5$ is shown.}
    \label{table:Result1}
    \resizebox{\textwidth}{!}{\begin{tabular}{|c|c|c|c|c|c|c|c|c|c|c|} \hline
    
        \textbf{Dataset} & \textbf{Approaches} & \multicolumn{3}{c|}{\textbf{ROUGE-1}} & \multicolumn{3}{c|} {\textbf{ROUGE-2}} & \multicolumn{3}{c|}{\textbf{ROUGE-L}}  \\ \cline{3-11}
        
        &  & \textbf{Precision} & \textbf{Recall} & \textbf{F1-score} & \textbf{Precision} & \textbf{Recall} & \textbf{F1-score} & \textbf{Precision} & \textbf{Recall} & \textbf{F1-score} \\ \hline

                 & $OntoDSumm$   & 0.55 & 0.53 & {\bf0.54} & 0.24 & 0.21 & 0.23 & 0.30 & 0.28 & 0.29 \\ \cline{2-11}
        ${D_1}$  & $B_1$         & 0.49 & {\bf0.58} & 0.53 & 0.24 & {\bf0.29} & {\bf0.26} & 0.31 & {\bf0.36} & {\bf0.33}\\ \cline{2-11}
                 & $B_2$         & {\bf 0.59} & 0.47 & 0.52 & {\bf 0.25} & 0.20 & 0.22 & {\bf0.32} & 0.27 & 0.29\\ \cline{2-11}
                 & $B_3$         & 0.44 & 0.53 & 0.48 & 0.18 & 0.22 & 0.20 & 0.25 & 0.29 & 0.27 \\ \cline{2-11}
                 & $B_4$         & 0.47 & 0.52 & 0.49 & 0.19 & 0.21 & 0.20 & 0.28 & 0.30 & 0.29 \\ \cline{1-11} \cline{1-11}
           
                 & $OntoDSumm$ & {\bf0.47} & {\bf0.45} & {\bf0.46} & {\bf 0.18} & {\bf0.18} & {\bf0.18} & {\bf 0.27} & {\bf0.27} & {\bf0.27}  \\ \cline{2-11}
        ${D_2}$  & $B_1$         & 0.32 & 0.35 & 0.33 & 0.14 & 0.15 & 0.14 & 0.23 & 0.25 & 0.24  \\ \cline{2-11}
                 & $B_2$         & 0.44 & 0.32 & 0.37 & 0.18 & 0.14 & 0.16 & 0.28 & 0.22 & 0.25  \\ \cline{2-11}
                 & $B_3$         & 0.39 & 0.44 & 0.41 & 0.11 & 0.13 & 0.12 & 0.23 & 0.26 & 0.24  \\ \cline{2-11}
                 & $B_4$         & 0.37 & 0.39 & 0.38 & 0.13 & 0.13 & 0.13 & 0.25 & 0.26 & 0.25  \\ \cline{1-11} \cline{1-11}

                 & $OntoDSumm$ & {\bf0.48} & {\bf0.45} & {\bf0.47} & {\bf0.20} & {\bf0.19} & {\bf0.20} & {\bf0.32} & {\bf0.29} & {\bf0.30} \\ \cline{2-11}
        ${D_3}$  & $B_1$         & 0.36 & 0.36 & 0.36 & 0.14 & 0.14 & 0.14 & 0.29 & 0.29 & 0.29  \\ \cline{2-11}
                 & $B_2$         & 0.50 & 0.34 & 0.40 & 0.23 & 0.15 & 0.18 & 0.32 & 0.23 & 0.27  \\ \cline{2-11}
                 & $B_3$         & 0.42 & 0.47 & 0.44 & 0.17 & 0.18 & 0.17 & 0.23 & 0.25 & 0.24  \\ \cline{2-11}
                 & $B_4$         & 0.41 & 0.40 & 0.41 & 0.15 & 0.15 & 0.15 & 0.29 & 0.29 & 0.29  \\ \cline{1-11} \cline{1-11}

                 & $OntoDSumm$ & {\bf0.48} & {\bf0.48} & {\bf0.47} & {\bf0.20} & {\bf0.19} & {\bf0.19} & {\bf0.29} & {\bf0.28} & {\bf0.28}  \\ \cline{2-11}
        ${D_4}$  & $B_1$         & 0.29 & 0.36 & 0.32 & 0.12 & 0.14 & 0.13 & 0.22 & 0.27 & 0.24  \\ \cline{2-11}
                 & $B_2$         & 0.48 & 0.35 & 0.41 & 0.19 & 0.14 & 0.16 & 0.29 & 0.23 & 0.26  \\ \cline{2-11}
                 & $B_3$         & 0.43 & 0.48 & 0.45 & 0.16 & 0.18 & 0.17 & 0.24 & 0.26 & 0.25  \\ \cline{2-11}
                 & $B_4$         & 0.37 & 0.35 & 0.36 & 0.14 & 0.13 & 0.13 & 0.28 & 0.26 & 0.27  \\ \cline{1-11} \cline{1-11}

                 & $OntoDSumm$ & {\bf0.59} & 0.59 & {\bf0.59} & {\bf0.35} & {\bf0.35} & {\bf0.35} & {\bf0.35} & 0.35 & {\bf0.35} \\ \cline{2-11}
        ${D_5}$  & $B_1$         & 0.35 & {\bf0.62} & 0.44 & 0.18 & 0.32 & 0.23 & 0.24 & {\bf 0.40} & 0.29 \\ \cline{2-11}
                 & $B_2$         & 0.48 & 0.54 & 0.51 & 0.22 & 0.25 & 0.23 & 0.26 & 0.29 & 0.28  \\ \cline{2-11}
                 & $B_3$         & 0.48 & 0.58 & 0.53 & 0.23 & 0.28 & 0.25 & 0.25 & 0.29 & 0.27  \\ \cline{2-11}
                 & $B_4$         & 0.39 & 0.38 & 0.38 & 0.16 & 0.16 & 0.16 & 0.32 & 0.32 & 0.32  \\ \cline{1-11} \cline{1-11}
    \end{tabular} }
\end{table*}

\subsection{Baselines} \label{s:base} 

\par We compare \textit{OntoDSumm} with the following state-of-the-art summarization approaches: 
\begin{enumerate}
    
    \item \textit{$B_1$}: Rudra et al.~\cite{rudra2019summarizing} proposed a summarization framework where they initially create a graph where the nodes are the most important disaster-specific keywords, and the edges represent the bigram relationship between a pair of keywords. Finally, they select the tweets in summary which can ensure maximum information coverage of the graph. 
    
    \item \textit{$B_2$}: Dutta et al.~\cite{dutta2018ensemble} propose an ensemble graph based summarization approach which initially generates summary by $9$ existing text summarization algorithms. The authors create a tweet similarity graph where the nodes represent the tweets present in the summary of any of the $9$ existing text summarization algorithms and the edges represent their content and context similarity. Finally, the authors follow a community detection algorithm to automatically identify the categories and then, select representative tweets from each category on the basis of length, informativeness and centrality scores to create the summary.
    
    \item \textit{$B_3$}: Rudra et al.~\cite{rudra2018identifying} propose a sub-event based summarization approach that initially identifies the sub-events and then, they generate a summary by selecting representative tweets by Integer Linear Programming based selection.
    
    \item \textit{$B_4$}: Nguyen et al.~\cite{nguyen2022towards} propose an abstractive summarization approach for disaster tweet summarization. \cite{nguyen2022towards} utilize a pre-trained BERT model to identify key-phrases from tweets and further, select those phrases which provide maximum information to generate the summary. For our experiments, we select those tweets in the summary which provide maximum coverage of key-phrases in the final summary.
    
    
    
\end{enumerate}

\subsection{Comparison with Existing Research Works}  \label{s:res}
\par We compare the summary generated by the \textit{OntoDSumm} and the existing research works with the ground truth summary on the basis of ROUGE-N score~\cite{lin2004rouge}. ROUGE-N score computes the overlapping words in the generated summary with a set of ground truth summary. We calculate precision, recall and F1-score for $3$ different variants of ROUGE-N score, i.e., N=$1$, $2$ and L, respectively. Our observations as shown in Table~\ref{table:Result1} and Table~\ref{table:Result2} indicate that \textit{OntoDSumm} ensure better ROUGE-N precision, recall and F1-scores in comparison with baselines. The improvement in summary scores of ROUGE-1 F1-score ranges from $1.85$\% to $65.46$\%, ROUGE-2 F1-score ranges from $4.35$\% to $76.47$\%  and ROUGE-L F1-score ranges from $2.63$\% to $25.00$\% respectively. The improvement is highest over the $B_1$ baseline and lowest with the $B_2$ baseline. The performance of the \textit{OntoDSumm} is the best for $D_6$ with $0.61-0.23$ and worst for $D_2$ with $0.47-0.18$ in Rouge-1, Rouge-2 and Rouge-L F1-scores.

\begin{table*}[ht]
    \centering 
    \caption{Precision, recall and F1-score of ROUGE-1, ROUGE-2 and ROUGE-L score of \textit{OntoDSumm} and baselines on $5$ datasets, i.e., $D_6-D_{10}$ is shown.}
    \label{table:Result2}
    \resizebox{\textwidth}{!}{\begin{tabular}{|c|c|c|c|c|c|c|c|c|c|c|} \hline
    
        \textbf{Dataset} & \textbf{Approaches} & \multicolumn{3}{c|}{\textbf{ROUGE-1}} & \multicolumn{3}{c|} {\textbf{ROUGE-2}} & \multicolumn{3}{c|}{\textbf{ROUGE-L}}  \\ \cline{3-11}
        
        &  & \textbf{Precision} & \textbf{Recall} & \textbf{F1-score} & \textbf{Precision} & \textbf{Recall} & \textbf{F1-score} & \textbf{Precision} & \textbf{Recall} & \textbf{F1-score} \\ \hline

                 & $OntoDSumm$ & {\bf0.61} & {\bf0.55} & {\bf0.57} & {\bf0.25} & {\bf0.23} & {\bf0.24} & {\bf0.31} & {\bf0.28} & {\bf0.30}  \\ \cline{2-11}
        ${D_6}$  & $B_1$         & 0.50 & 0.47 & 0.49 & 0.23 & 0.22 & 0.22 & 0.29 & 0.28 & 0.29  \\ \cline{2-11}
                 & $B_2$         & 0.57 & 0.41 & 0.48 & 0.21 & 0.15 & 0.18 & 0.29 & 0.22 & 0.25  \\ \cline{2-11}
                 & $B_3$         & 0.50 & 0.55 & 0.52 & 0.20 & 0.22 & 0.21 & 0.25 & 0.27 & 0.23  \\ \cline{2-11}
                 & $B_4$         & 0.53 & 0.54 & 0.53 & 0.21 & 0.21 & 0.21 & 0.29 & 0.29 & 0.29 \\ \cline{1-11} \cline{1-11}

                 & $OntoDSumm$ & 0.53 & 0.49 & {\bf0.51} & {\bf0.18} & {\bf0.17} & {\bf0.17} & {\bf0.28} & {\bf0.24} & {\bf0.26} \\ \cline{2-11}
        ${D_7}$  & $B_1$         & 0.45 & {\bf0.51} & 0.48 & 0.12 & 0.14 & 0.13 & 0.24 & 0.21 & 0.22  \\ \cline{2-11}
                 & $B_2$         & {\bf0.57} & 0.40 & 0.47 & 0.17 & 0.12 & 0.14 & 0.27 & 0.21 & 0.22  \\ \cline{2-11}
                 & $B_3$         & 0.49 & 0.40 & 0.44 & 0.14 & 0.11 & 0.12 & 0.24 & 0.20 & 0.22  \\ \cline{2-11}
                 & $B_4$         & 0.49 & 0.51 & 0.50 & 0.16 & 0.16 & 0.16 & 0.25 & 0.24 & 0.25  \\ \cline{1-11} \cline{1-11}

                 & $OntoDSumm$ & {\bf0.53} & {\bf0.50} & {\bf0.52} & {\bf0.19} & {\bf0.18} & {\bf0.19} & {\bf0.27} & {\bf0.26} & {\bf0.27}  \\ \cline{2-11}
        ${D_8}$  & $B_1$         & 0.43 & 0.48 & 0.45 & 0.13 & 0.14 & 0.13 & 0.22 & 0.24 & 0.23  \\ \cline{2-11}
                 & $B_2$         & 0.51 & 0.43 & 0.46 & 0.15 & 0.13 & 0.14 & 0.26 & 0.22 & 0.24  \\ \cline{2-11}
                 & $B_3$         & 0.43 & 0.46 & 0.44 & 0.13 & 0.14 & 0.14 & 0.22 & 0.24 & 0.23  \\ \cline{2-11}
                 & $B_4$         & 0.51 & 0.49 & 0.50 & 0.18 & 0.16 & 0.17 & 0.26 & 0.24 & 0.25  \\ \cline{1-11} \cline{1-11}   

                 & $OntoDSumm$ & {\bf0.55} & {\bf0.53} & {\bf0.54} & {\bf0.18} & {\bf0.17} & {\bf0.17} & {\bf0.25} & {\bf0.24} & {\bf0.24}  \\ \cline{2-11}
        ${D_9}$  & $B_1$         & 0.26 & 0.16 & 0.20 & 0.05 & 0.03 & 0.04 & 0.25 & 0.17 & 0.20  \\ \cline{2-11}
                 & $B_2$         & 0.50 & 0.44 & 0.47 & 0.15 & 0.13 & 0.14 & 0.22 & 0.20 & 0.21  \\ \cline{2-11}
                 & $B_3$         & 0.45 & 0.45 & 0.45 & 0.11 & 0.11 & 0.11 & 0.21 & 0.21 & 0.21  \\ \cline{2-11}
                 & $B_4$         & 0.44 & 0.43 & 0.44 & 0.11 & 0.10 & 0.10 & 0.21 & 0.20 & 0.20  \\ \cline{1-11} \cline{1-11}
                       
                 & $OntoDSumm$ & {\bf0.56} & {\bf0.55} & {\bf0.55} & {\bf0.18} & {\bf0.17} & {\bf0.17} & {\bf0.24} & {\bf0.23} & {\bf0.24}  \\ \cline{2-11}
     ${D_{10}}$  & $B_1$         & 0.26 & 0.15 & 0.19 & 0.06 & 0.03 & 0.04 & 0.22 & 0.15 & 0.18  \\ \cline{2-11}
                 & $B_2$         & 0.51 & 0.45 & 0.48 & 0.10 & 0.09 & 0.10 & 0.22 & 0.19 & 0.20  \\ \cline{2-11}
                 & $B_3$         & 0.50 & 0.49 & 0.50 & 0.12 & 0.12 & 0.12 & 0.22 & 0.22 & 0.22  \\ \cline{2-11}
                 & $B_4$         & 0.52 & 0.51 & 0.52 & 0.12 & 0.12 & 0.12 & 0.21 & 0.20 & 0.21  \\ \cline{1-11} \cline{1-11}
    \end{tabular} }
\end{table*}

\subsection{Ablation Experiments}\label{s:ablexp}
\par To understand the effectiveness of each Phase of \textit{OntoDSumm}, we compare and validate the performance of \textit{OntoDSumm} with its different variants and study the role of ontology based summarization with the identification of category importance.   

\subsubsection{Phase-I}\label{s:ablphase1}

\par In this experiment, we compare the proposed Phase-I with the following variant of the Phase-I of \textit{OntoDSumm}:

\begin{enumerate}
    \item In \textbf{OntoDSumm-1A}~\footnote{In the name of a variant of   \textit{OntoDSumm}, i.e., \textit{OntoDSumm-1A}, the first number represents the Phase, and the second letter represents the sequence of the ablation study in that Phase.}, we use the original keywords provided in  \textit{Empathi} and do not integrate our proposed extended vocabulary for each category to identify the category of a tweet. 
\end{enumerate}

\subsubsection{Phase-II}\label{s:ablphase2}
\par To understand the different aspects of Phase-II and its role in the performance of \textit{OntoDSumm}, we generate $3$ different variants of Phase-II and compare their performance with \textit{OntoDSumm} which are as follows : 

\begin{enumerate}
    \item In \textbf{OntoDSumm-2A}, we assume each category is given equal importance, i.e., equal number of tweets are selected from each category.
    \item In \textbf{OntoDSumm-2B}, we use Ridge Regression model to predict the number of tweets to be selected from a category. 
    \item In \textbf{OntoDSumm-2C}, we use Bayesian Regression model to predict the number of tweets to be selected from a category. 
\end{enumerate}

\subsubsection{Phase-III}\label{s:ablphase3}
\par To validate the performance and specificity of \textit{DMMR} for summarization with respect to disasters, we compare the proposed \textit{DMMR} with its different variants, which are as follows:

\begin{enumerate}
    \item In \textbf{OntoDSumm-3A}, we select that tweet in each iteration which provides the maximum information coverage of a category, i.e., has maximum Semantic Similarity score (as discussed in Section~\ref{s:phase1}) with the category keywords into the summary and therefore, we do not consider diversity while selection of the tweet. 
    
    \item In \textbf{OntoDSumm-3B}, we use K-means clustering algorithm~\cite{macqueen1967some} to select the K tweets from a category into the summary. This K is chosen depending on importance of a category. We select the centroid of each K-cluster as the representative tweets of the category.
    
    \item In \textbf{OntoDSumm-3C}, we select that tweet in each iteration from a category which has the maximum Eigenvector centrality score~\cite{bonacich1972factoring} into the summary. 
    
    \item In \textbf{OntoDSumm-3D}, we select that tweet in each iteration from a category  which has the maximum PageRank score~\cite{brin1998anatomy} into the summary. 
    
    \item In \textbf{OntoDSumm-3E}, we select that tweet in each iteration from a category which has the maximum Maximal Marginal Relevance based (MMR) score~\cite{carbonell1998use} into the summary. 
\end{enumerate}

\subsubsection{Results and Discussions}\label{s:result}

\par
\begin{figure*}[t!]
    \includegraphics[width=1.0\textwidth]{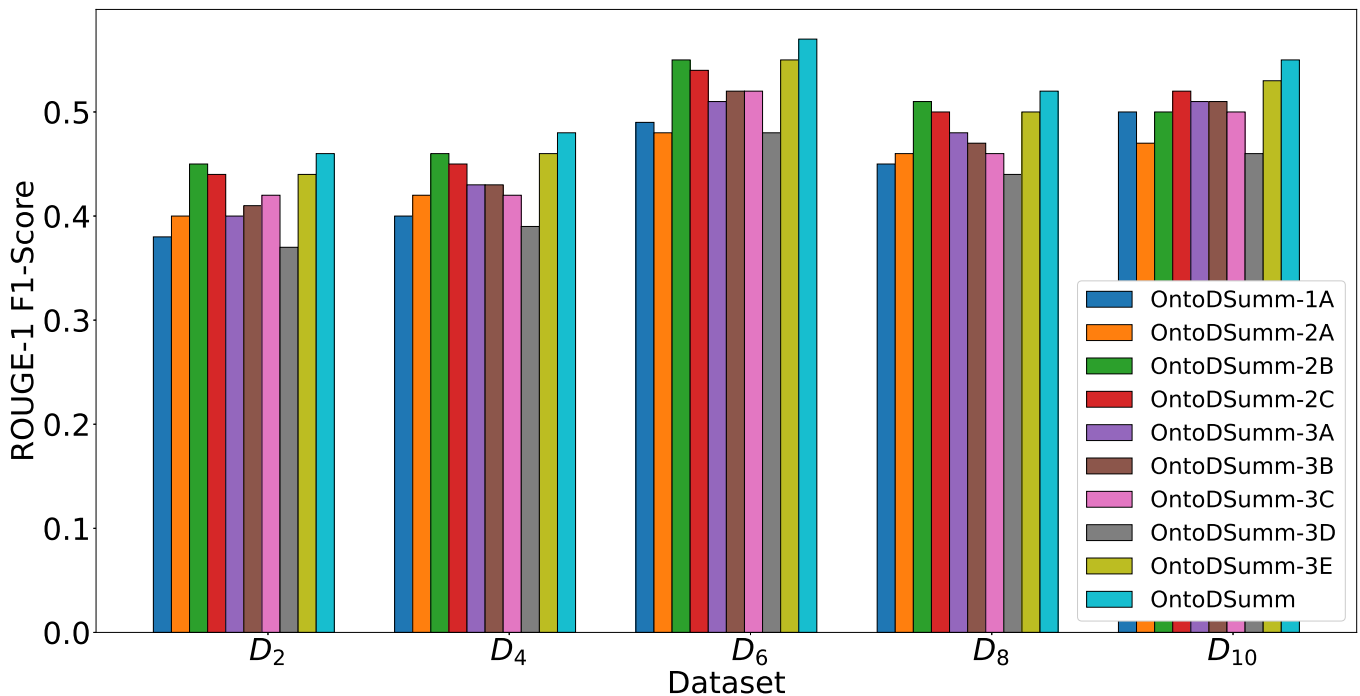}
\caption{We show F1-score of ROUGE-1 for $9$ different variants of our proposed \textit{OntoDSumm}, i.e., \textit{OntoDSumm-1A}, \textit{OntoDSumm-2A}, \textit{OntoDSumm-2B}, \textit{OntoDSumm-2C}, \textit{OntoDSumm-3A}, \textit{OntoDSumm-3B}, \textit{OntoDSumm-3C}, \textit{OntoDSumm-3D}, and \textit{OntoDSumm-3E} with  \textit{OntoDSumm} on $5$ different disasters, such as \textit{Uttarakhand Flood} ($D_2$), \textit{Hyderabad Blast} ($D_4$), \textit{ Los Angeles International Airport Shooting} ($D_6$), \textit{Puebla Mexico Earthquake} ($D_8$), and \textit{Midwestern U.S. Floods} ($D_{10}$).}
\label{fig:ablationFig}
\end{figure*}

 We now compare and validate the performance of \textit{OntoDSumm} with the discussed different variants on $5$ different disasters, such as $D_2$, $D_4$, $D_6$, $D_8$ and $D_{10}$ on the basis of ROUGE-1 F1-score. Our observations as shown in Figure~\ref{fig:ablationFig} indicate that \textit{OntoDSumm} is better by around $2-20\%$ in ROUGE-1 F1-score across the disasters.

\begin{itemize}
    \item Phase-I: We observe that \textit{OntoDSumm} outperforms \textit{OntoDSumm-1A}, which uses only the original keywords provided in \textit{Empathi} by $9.09-17.39\%$ in ROUGE-1 F1-score which highlights the effectiveness of keyword extension using Wikipedia pages. 
    \item Phase-II: We observe \textit{OntoDSumm} outperforms \textit{OntoDSumm-2A}, which gives the equal importance to each category by $11.53-15.78\%$ in ROUGE-1 F1-score which highlights the need to consider the specific importance of each category corresponding to a disaster. Additionally, we validate the decision to use Linear Regression over the Bayesian Regression or Ridge Regression as \textit{OntoDSumm} outperforms \textit{OntoDSumm-2C} and \textit{OntoDSumm-2B} by $1.92-9.09\%$ in ROUGE-1 F1-score. 
    \item Phase-III: Finally, we observe that the inclusion of \textit{DMMR} in \textit{OntoDSumm} provides an increase of $3.63-19.56\%$ in ROUGE-1 F1-score when compared to different variants of Phase-III, i.e, \textit{OntoDSumm-3A}, \textit{OntoDSumm-3B}, \textit{OntoDSumm-3C}, \textit{OntoDSumm-3D}, and \textit{OntoDSumm-3E}. This justifies that domain knowledge is immensely helpful even in finding out important tweets from each category. Additionally, we observe that the performance gain of \textit{OntoDSumm} is highest over the \textit{OntoDSumm-3D}, which uses PageRank as Pagerank does not consider the diversity criteria, while it is lowest against \textit{OntoDSumm-3E}, which uses the MMR for selecting the tweets from each category. We also observe that inclusion of the novel \textit{DMMR} for Phase-III provides the maximum increase in performance of \textit{OntoDSumm}.
\end{itemize}


\par 

\subsection{Identification of Category of a Tweet}\label{s:catidenti}
\par In this Subsection, we evaluate the effectiveness of Phase-I, i.e., the proposed unsupervised tweet category identification with an existing unsupervised approach used in~\cite{dutta2019community}. We select the approach followed by Dutta et al.~\cite{dutta2019community} where the authors utilize Louvain community detection algorithm~\cite{traag2015faster} to identify communities which inherently represents categories related to the disaster. For our experiments, we select $20\%$ of the total tweets randomly for a disaster event. We, then, asked $3$ annotators to manually determine the category of selected tweets. We assign final category level based on majority voting. We, then, compare the identified category by \textit{OntoDSumm} and the approach used in~\cite{dutta2019community} respectively. We repeat this for $4$ disaster events. We found that F1-Score by \textit{OntoDSumm} and the approach used in~\cite{dutta2019community} are in the rage of $97.8-99.7\%$ and $53.3-74.9\%$ respectively, as shown in Table~\ref{table:catdet}. Therefore, on the basis of our experiments, we can confirm that utilization of the domain knowledge of ontology ensures high efficiency in tweet categorization.

\begin{table}
    \centering 
     \caption{We show F1-score of Phase-I of \textit{OntoDSumm} and unsupervised tweet category identification approach used in~\cite{dutta2019community} on comparing the volunteers annotations for $4$ disasters, such as, $D_3$, $D_6$, $D_8$, and $D_{10}$.} 
    \label{table:catdet}
    \begin{tabular}{|c|c|c|} \hline
        \textbf{Dataset} & \textbf{Approach} & \textbf{F1-score} \\ \hline

        ${D_3}$     & Phase-I of \textit{OntoDSumm}                 & \bf{0.978} \\ \cline{2-3} 
                    & Approach used in~\cite{dutta2019community}    & 0.653      \\ \hline
                    
        ${D_6}$     & Phase-I of \textit{OntoDSumm}                 & \bf{0.995} \\ \cline{2-3} 
                    & Approach used in~\cite{dutta2019community}    & 0.533      \\ \hline
                    
        ${D_8}$     & Phase-I of \textit{OntoDSumm}                 & \bf{0.984} \\ \cline{2-3} 
                    & Approach used in~\cite{dutta2019community}    & 0.571      \\ \hline
        
        ${D_{10}}$  & Phase-I of \textit{OntoDSumm}                 & \bf{0.997} \\ \cline{2-3} 
                    & Approach used in~\cite{dutta2019community}    &  0.749     \\ \hline
    \end{tabular} 
\end{table}

\subsection{Understanding Disaster Similarity Index} \label{s:similar}

\par On comparing $DisSIM(D_x,D_y)$ for every disaster pair, we observe that any disaster (say $D_x$) having maximum similarity score with another disaster (say $D_y$) actually belongs to the same continent (such as USA or Asia) and is of the same type (i.e., man-made or natural). For example,  \textit{Uttarakhand Flood} ($D_2$) has the maximum similarity score with disaster \textit{Pakistan Earthquake} ($D_9$) among all the disasters (both the disasters belong to Asia and type is natural). Similarly, \textit{Sandy Hook Elementary School Shooting} ($D_1$) has the maximum similarity score with \textit{Los Angeles International Airport Shooting} ($D_6$) among all the disasters (both disasters belong to USA and man-made). Therefore, we term $D_x$ and $D_y$ as \textit{homogeneous disasters} if they belong to the same continent and is of the same type. Subsequently, we term $D_x$ and $D_y$ as \textit{heterogeneous disasters} if they do not satisfy either of these two conditions. We validate next the impact of training on a \textit{homogeneous disaster} or a \textit{heterogeneous disaster} on the summary quality. We show the results in Table~\ref{table:Combined-similarity}. 

\par For each disaster $D_x$, we predict the number of tweets to be selected from each category as predicted by the linear regression model if it was trained on \textit{homogeneous disaster} or \textit{heterogeneous disaster}. We, then, select the representative tweets from each category based on the predicted number of tweets to create the summary and finally, compare the generated summary with the ground-truth summary on the basis of ROUGE-N scores. For the experiments, we randomly select $6$ disasters, such as, $D_1$, $D_2$, $D_5$, $D_7$, $D_8$, and $D_{10}$. We show the precision, recall, and F1-score for $3$ different variants of the ROUGE-N score, i.e., N=1, 2, and L, respectively, in Table~\ref{table:cat_train}. Our observations indicate that there is around $34-43\%$ and $8-30\%$ increase in ROUGE-2 F1-score for any disaster in Asia and USA, respectively, for both man-made and natural disasters. Therefore, our observations indicate that an effective summary for $D_x$ can be ensured if we identify the importance of categories from a $D_y$ is which of the same type and from the same continent as $D_y$.

\begin{table*}[ht]
    \centering 
    \caption{We show the \textit{Disaster Similarity Index} for every disaster pair. The first row and first column of this Table represent the various disasters. }
    \label{table:Combined-similarity} 
    \begin{tabular} {|p{0.04\linewidth}|p{0.05\linewidth}|p{0.05\linewidth}|p{0.05\linewidth}|p{0.05\linewidth}|p{0.05\linewidth}|p{0.05\linewidth}|p{0.05\linewidth}|p{0.05\linewidth}|p{0.05\linewidth}|p{0.05\linewidth}|} \hline
        {} & \textbf{$D_1$} & \textbf{$D_2$} & \textbf{$D_3$} & \textbf{$D_4$} & \textbf{$D_5$} & \textbf{$D_6$} & \textbf{$D_7$} & \textbf{$D_8$} & \textbf{$D_9$} & \textbf{$D_{10}$}\\ \hline 
        
        $D_1$    & 	        & 0.39    & 0.38  & 0.39    & 0.36    &\bf{0.53}& 0.40    & 0.26    & 0.40    & 0.24  \\\hline
        $D_2$    & 0.39     &         & 0.50  & 0.46    & 0.44    & 0.37    & 0.43    & 0.43    &\bf{0.52}& 0.42 \\\hline
        $D_3$    & 0.38	    & 0.50    & 	  & 0.40    & 0.40	  & 0.40    & 0.44    & 0.36    &\bf{0.51}& 0.34 \\\hline
        $D_4$    & 0.39	    & 0.46    & 0.40  & 	    &\bf{0.53}& 0.44    & 0.42    & 0.32    & 0.38    & 0.34 \\\hline
        $D_5$    & 0.36	    & 0.44    & 0.40  &\bf{0.53}&         & 0.41    & 0.35    & 0.36    & 0.38    & 0.29 \\\hline
        $D_6$    &\bf{0.53} & 0.37    & 0.40  & 0.44    & 0.41    &         & 0.36    & 0.25    & 0.37    & 0.28 \\\hline
        $D_7$    & 0.40	    & 0.43    & 0.44  & 0.42    & 0.35    & 0.36    &         &\bf{0.52}& 0.40    & 0.50 \\\hline
        $D_8$    & 0.26     & 0.43    & 0.36  & 0.32    & 0.36    & 0.25    &\bf{0.52}&         & 0.42    & 0.52 \\\hline
        $D_9$    & 0.40	    &\bf{0.52}& 0.51  & 0.38    & 0.38    & 0.37    & 0.40    & 0.42    &         & 0.41 \\\hline
        $D_{10}$ & 0.24	    & 0.42    & 0.34  & 0.34    & 0.29    & 0.28    & 0.50    &\bf{0.52}& 0.41    & 	   \\\hline
    \end{tabular}
\end{table*}

\begin{table*}[ht]
    \centering 
    \caption{We show F1-score of ROUGE-1, ROUGE-2 and ROUGE-L of \textit{OntoDSumm} for training a linear regression model on a dataset from homogeneous and heterogeneous disasters on $6$ datasets, such as $D_1$, $D_2$, $D_5$, $D_7$, $D_8$, and $D_{10}$.}
    \label{table:cat_train}
    \begin{tabular}{|c|c|c|c|c|c|c|c|c|c|} \hline
        \textbf{Dataset} & \textbf{Training} & \textbf{ROUGE-1} & \textbf{ROUGE-2} & \textbf{ROUGE-L} & \textbf{Dataset} & \textbf{Training} & \textbf{ROUGE-1} & \textbf{ROUGE-2} & \textbf{ROUGE-L} \\ \cline{3-5} \cline{8-10}
        
        & \textbf{disaster} & \textbf{F1-score} & \textbf{F1-score} & \textbf{F1-score} & & \textbf{disaster} & \textbf{F1-score} & \textbf{F1-score} & \textbf{F1-score} \\ \hline

                                &           {$D_5$}        & 0.45 & 0.18 & 0.24 &                                       & {$D_1$} & 0.48 & 0.11 & 0.23 \\ \cline{2-5} \cline{7-10}
    \scalebox{1.1}{\bm{${D_1}$}}&\scalebox{1.1}{\bm{$D_6$}}&\bf0.54&\bf0.23&\bf0.29&    \scalebox{1.1}{\bm{${D_7}$}}     & {$D_5$} & 0.42 & 0.12 & 0.21 \\ \cline{2-5} \cline{7-10}
                                &           {$D_9$}        & 0.49 & 0.21 & 0.25 &                      &\scalebox{1.1}{\bm{$D_8$}}&\bf0.51&\bf0.17&\bf0.26 \\ \cline{2-5} \cline{7-10}
                                &           $D_{10}$       & 0.42 & 0.16 & 0.24 &                                       & {$D_9$} & 0.46 & 0.12 & 0.22 \\ \hline

                                &           {$D_1$}        & 0.39 & 0.12 & 0.21 &                                       & {$D_1$} & 0.45 & 0.14 & 0.23 \\ \cline{2-5} \cline{7-10}
    \scalebox{1.1}{\bm{${D_2}$}}&           {$D_5$}        & 0.41 & 0.12 & 0.22 &          \scalebox{1.1}{\bm{${D_8}$}} & {$D_2$} & 0.43 & 0.10 & 0.20 \\ \cline{2-5} \cline{7-10}
                                &           {$D_8$}        & 0.39 & 0.12 & 0.22 &                                       & {$D_4$} & 0.46 & 0.14 & 0.22 \\ \cline{2-5} \cline{7-10}
                                &\scalebox{1.1}{\bm{$D_9$}}&\bf0.46&\bf0.18&\bf0.27&                  & \scalebox{1.1}{\bm{$D_7$}}&\bf0.52&\bf0.19&\bf0.27 \\ \hline   
                 
                                &\scalebox{1.1}{\bm{$D_4$}}&\bf0.59&\bf0.35&\bf0.35&                                    & {$D_2$} & 0.48 & 0.10 & 0.19 \\ \cline{2-5} \cline{7-10}
    \scalebox{1.1}{\bm{${D_5}$}}&           {$D_6$}        & 0.48 & 0.23 & 0.26 &       \scalebox{1.1}{\bm{${D_{10}}$}} & {$D_4$} & 0.50 & 0.13 & 0.20 \\ \cline{2-5} \cline{7-10}
                                &           {$D_7$}        & 0.49 & 0.22 & 0.29 &                                       & {$D_6$} & 0.44 & 0.08 & 0.18 \\ \cline{2-5} \cline{7-10}
                                &           {$D_9$}        & 0.46 & 0.20 & 0.26 &                     &\scalebox{1.1}{\bm{$D_8$}}& \bf0.55 & \bf0.17 & \bf0.24 \\ \hline
    \end{tabular} 
\end{table*}

\subsection{Limitations of \textit{OntoDSumm}} \label{s:fail}

\par We discuss the limitations of \textit{OntoDSumm} that we have observed next.  

\begin{enumerate}
    \item \textit{Dependence on Existing Ontology} :  As \textit{OntoDSumm} relies on an existing ontology to identify the categories of a tweet, it is not directly applicable for other summarization applications, like news events, user opinions regarding products, etc., unless an existing ontology for that application is available. We are working towards developing an ontology automatically from publicly available resources for any application as a future direction so that \textit{OntoDSumm} is not dependent on an existing ontology.
    
    \item \textit{Identification of category of a Tweet} : 
    \textit{OntoDSumm} can not identify the categories of all tweets. For example, we found among all the disasters, \textit{OntoDSumm} performs the worst for $D_4$, where it could not identify the category for around $19.10-31.59\%$ of the tweets. Therefore, we believe Phase-I of \textit{OntoDSumm} could be further improved such that the category of more number of tweets could be effectively identified. 
    
    
    
    \item \textit{Limitation of Disaster Dataset} : In Phase-II of \textit{OntoDSumm}, we find most similar disaster for finding out the category importance of the given disaster. However, it is quite possible that this Phase will perform poorly when existing dataset does not result in high similarity score. To overcome this a more diverse and more enriched disaster dataset is required. 
\end{enumerate}

\section{Conclusions and Future works} \label{s:con}

\par In this paper, we propose \textit{OntoDSumm} which can generate a tweet summary for a disaster with minimal human intervention. However, summarization of tweets related to a disaster has several challenges, like, automatic identification of the category of a tweet, determination of representativeness of each category in summary with respect to the disaster and ensure information coverage of each category while maintaining diversity in the summary. In order to handle these challenges, we propose a three-phase approach which can handle each challenge systematically with high effectiveness and minimal human intervention. We believe the incorporation of domain knowledge through ontology for category identification, automatic understanding and knowledge transfer across different disasters to gauge the importance of a category and finally, a selection mechanism specifically designed for disasters ensures the high performance of \textit{OntoDSumm}. Our experimental analysis shows that \textit{OntoDSumm} can ensure $2-77\%$ increase in Rouge-N F1-scores over existing research works. Additionally, we validate experimentally each Phase of \textit{OntoDSumm} for generating a summary. 
As a future direction, we are working towards mitigation of all the limitations mentioned in Section~\ref{s:fail}.  

\bibliographystyle{IEEEtran}
\bibliography{journal_bib.bib}

\end{document}